\documentclass[letterpaper,12pt]{article}
\pdfoutput=1

\usepackage{jheppub}

\usepackage{caption}

\DeclareMathOperator{\Gr}{Gr}

\title{
\Large A Note on One-loop Cluster Adjacency in $\mathcal{N}=4$ SYM
}

\author[1]{Jorge Mago,}\emailAdd{jorge\_mago@brown.edu}
\author[1]{Anders Schreiber,}\emailAdd{anders\_schreiber@brown.edu}
\author[1,2]{Marcus Spradlin}\emailAdd{marcus\_spradlin@brown.edu}
\author[1]{and Anastasia Volovich}\emailAdd{anastasia\_volovich@brown.edu}

\affiliation[1]{Department of Physics,
Brown University,
Providence, RI 02912, USA}

\affiliation[2]{Brown Theoretical Physics Center,
Brown University,
Providence, RI 02912, USA}

\abstract{We study cluster adjacency conjectures for amplitudes in maximally supersymmetric Yang-Mills theory. We show that the $n$-point one-loop NMHV ratio function satisfies Steinmann cluster adjacency. We also show that the one-loop BDS-like normalized NMHV amplitude satisfies cluster adjacency between Yangian invariants and final symbol entries up to 9-points. We present conjectures for cluster adjacency properties of Pl{\"u}cker coordinates, quadratic cluster variables, and NMHV Yangian invariants that generalize the notion of weak separation.
}

\begin{document}

\maketitle


\section{Introduction}

Cluster algebras \cite{Fomin:2001i,FWZ1,FWZ2} of Grassmannian type \cite{Gekhtman:2003, Scott} have been found to play a significant role in the mathematical structure of scattering amplitudes in planar maximally supersymmetric Yang-Mills theory ($\mathcal{N}=4$ SYM)  \cite{Golden:2013xva, Golden:2014xqa}, constraining the structure of amplitudes at the level of symbols and cobrackets \cite{Golden:2014pua, Golden:2014xqa, Harrington:2015bdt, Golden:2018gtk}. The recently introduced cluster adjacency principle \cite{Drummond:2017ssj} has opened a new line of research in this topic, shedding light on even deeper connections between amplitudes and cluster algebras. This principle applies conjecturally to various aspects of the analytic structure of amplitudes in $\mathcal{N}=4$ SYM. The many guises of cluster adjacency, at the level of symbols \cite{Golden:2019kks}, Yangian invariants \cite{Mago:2019waa,Lukowski:2019sxw}, and the correlation between them \cite{Drummond:2018dfd} have also been exploited to help compute new amplitudes via bootstrap \cite{Drummond:2018caf}. These mathematical properties however are perhaps somewhat obscure, and although it is understood that cluster adjacency of a symbol implies the Steinmann relations \cite{Drummond:2017ssj}, its other manifestations have less clear physical interpretations (see however~\cite{GP}, which establishes interesting new connections between cluster adjacency and Landau singularities).
Even finer notions of cluster adjacency, that more strictly constrain pairs of adjacent symbol letters, have recently been studied in~\cite{Henke:2019hve,Drummond:2020kqg}.

In this paper we show that the one-loop NMHV amplitudes in $\mathcal{N}=4$ SYM theory satisfy symbol-level cluster adjacency for all $n$, and we check that for $n=9$ the amplitude can be written in a form that exhibits adjacency between final symbol entries and R-invariants, supporting the conjectures
of \cite{Drummond:2017ssj, Drummond:2018dfd}.
The outline of this paper is as follows.
In Section 2 we review the kinematics of $\mathcal{N}=4$ SYM and the bracket test used to assess cluster adjacency. In Section 3 we review formulas for the amplitudes to which we apply the bracket test. In Section 4 we present our analysis and results as well as new cluster adjacency conjectures for Pl{\"u}cker coordinates and cluster variables that are quadratic in Pl\"uckers.
These conjectures generalize the notion of weak separation~\cite{LZ,Oh:2011}.

\section{Cluster Adjacency and the Sklyanin Bracket}
\label{sec:two}

In $\mathcal{N} = 4 $ SYM the kinematics of scattering of $n$ massless particles is described by a collection of $n$ \emph{momentum twistors} \cite{Hodges:2009hk}, $Z_1^I , \ldots , Z_n^I$, each of which is a four-component ($I \in \{ 1, \ldots , 4 \}$) homogeneous coordinate on $\mathbb{P}^3$. Thanks to dual conformal symmetry~\cite{Drummond:2008vq} the collection of momentum twistors have a $GL(4)$ redundancy and thus can be taken to represent points in $\Gr(4,n)$.  By an appropriate choice of gauge we can take
\begin{align} \label{zmatrix}
Z =
\begin{pmatrix}
Z_{1}^1 & \cdots & Z_{n}^1 \\
Z_{1}^2 & \cdots & Z_{n}^2 \\
Z_{1}^3 & \cdots & Z_{n}^3 \\
Z_{1}^4 & \cdots & Z_{n}^4
\end{pmatrix} \underset{GL(4)}{\longrightarrow}
\begin{pmatrix}
1 & 0 & 0 & 0 & y^1{}_5 & \cdots & y^1{}_n \\
0 & 1 & 0 & 0 & y^2{}_5 & \cdots & y^2{}_n \\
0 & 0 & 1 & 0 & y^3{}_5 & \cdots & y^3{}_n \\
0 & 0 & 0 & 1 & y^4{}_5 & \cdots & y^4{}_n \\
\end{pmatrix}.
\end{align}
The degrees of freedom are given by $y^I{}_a = (-1)^I \langle \{ 1,2,3,4 \} \setminus \{ I \} , a \rangle / \langle 1 , 2 , 3, 4 \rangle$ for $a = 5, 6, \ldots , n$, with
\begin{align} \label{momfourbrack}
\langle a,b,c,d \rangle &\equiv \epsilon_{ijkl} Z_a^i Z_b^j Z_c^k Z_d^l
\end{align}
denoting Pl\"ucker coordinates on $\Gr(4,n)$.
Throughout this paper we will make use of the relation between momentum twistors and dual momenta~\cite{Drummond:2008vq}
\begin{align} \label{xfourbracket}
x_{ij}^2 = \frac{\langle i{-}1 \,,  i \,, j{-}1 \,, j \rangle 	}{\langle i{-}1 \, i \rangle \langle j{-}1 \, j \rangle} \, ,
\end{align}
where $\langle i \, j \rangle$ is the usual spinor helicity bracket (that completely drops out of our analysis due to cancellations guaranteed by dual conformal symmetry).

The fact that \eqref{momfourbrack} are cluster variables of the $\Gr(4,n)$ cluster algebra plays a constraining role in the analytic structure of amplitudes in $\mathcal{N}=4$ SYM through the notion of cluster adjacency \cite{Drummond:2017ssj} and it is therefore of interest to test the cluster adjacency properties of amplitudes. Two cluster variables are cluster adjacent if they appear together in a common cluster of the cluster algebra (this notion is also called ``cluster compatibility''). To test whether two given variables are cluster adjacent one can use the Poisson structure of the cluster algebra \cite{Gekhtman:2003}, which is related to the Sklyanin bracket \cite{Sklyanin:1982tf}. We call this the \emph{bracket test} and was first applied to amplitudes in \cite{Golden:2019kks}. In terms of the parameters of \eqref{zmatrix}, the Sklyanin bracket is given by
\begin{align}
\{ y^I{}_a, y^J{}_b \} = \frac{1}{2} (\text{sign} (J-I) - \text{sign} (b-a) ) y^J{}_a y^I{}_b \, ,
\end{align}
which extends to arbitrary functions as
\begin{align} \label{sklyaningen}
\{ f(y) , g(y) \} = \sum_{a, b = 5}^n \sum_{I, J =1}^4 \frac{\partial f}{\partial y^I{}_a} \frac{\partial g}{\partial y^J{}_b} \{ y^I{}_a, y^J{}_b \} \, .
\end{align}
The \emph{bracket test} is then the following conjecture: two cluster variables $a_i$ and $a_j$ are cluster adjacent iff
\begin{align} \label{sklyaninnorm}
\Omega_{ij} = \{ \log  a_i, \log a_j \} \in \frac{1}{2} \mathbb{Z} \, .
\end{align}
Note that whenever $i,j,k,l$ are cyclically adjacent,
$\langle i,j,k,l\rangle$ is a frozen variable and is therefore
automatically adjacent
with every cluster variable.

The aim of this paper is to  provide evidence for two cluster adjacency conjectures for loop amplitudes of generalized polylogarithm type \cite{Drummond:2017ssj}: \\

\noindent\emph{Conjecture 1 ``Steinmann cluster adjacency''}: Every pair of adjacent entries in the  symbol of an amplitude is cluster adjacent.\\

This type of cluster adjacency implies the extended
Steinmann relations at all particle multiplicities \cite{Golden:2019kks}.
In fact it appears to be equivalent to the extended Steinmann
conditions of \cite{Caron-Huot:2019bsq} for all known integrable
symbols with physical first entries (that means, of the form
$\langle i, i+1,j,j+1 \rangle$). \\

\noindent \emph{Conjecture 2 ``Final entry cluster adjacency''}: There exists a representation of the symbol of an amplitude in which the final symbol entry in every term is cluster adjacent to all poles of the Yangian invariant that term multiplies.\\

Support for these conjectures was given for NMHV amplitudes at 6- and 7-points in \cite{Drummond:2018caf,Drummond:2018dfd} (to all loop order at which these amplitudes are currently known), and for one- and two-loop MHV amplitudes (to which only the first conjecture applies) at all multipliticies in~\cite{Golden:2019kks}.

\section{One-loop Amplitudes}

To demonstrate the cluster adjacency of NMHV amplitudes with respect to the conjectures in Section \ref{sec:two} we need to work with appropriate finite quantities after IR divergences have been subtracted.
To this end we will be working with two types of regulators at one loop: BDS~\cite{Bern:2005iz} and BDS-like~\cite{Alday:2009dv} normalized amplitudes. In this section we review these regulators and the one-loop amplitudes relevant for our computations.

\subsection{BDS- and BDS-like Subtracted Amplitudes}

We start by reviewing the BDS normalized amplitude, which was first introduced in \cite{Bern:2005iz}. Consider the $n$-point MHV amplitude $\mathcal{A}_n^{\text{MHV}}$ in planar $\mathcal{N}=4$ SYM with gauge group $SU(N_c)$ coupling constant $g_{\text{YM}}$, where the tree-level amplitude has been factored out. Evaluating the amplitude in $4{-}2 \epsilon$ dimensions regulates the IR divergences.
The BDS normalization involves dividing all amplitudes by the factor
\begin{align} \label{bdsansatz}
A_n^{\text{BDS}} = \exp\left[ \sum_{L=1}^\infty g^{2L} \left( \frac{f^{(L)} (\epsilon)}{2} A_n^{(1)} (L \epsilon) + C^{(L)} \right) \right]  \, ,
\end{align}
that encapsulates all IR divergences.  Here
where $g^2 = \frac{g^2_{\text{YM} } N_c}{16 \pi^2}$ is the 't Hooft coupling, the superscript $(L)$ on any function denotes its $\mathcal{O}(g^{2L})$ term, $C^{(L)}$ is a transcendental constant and $f(\epsilon) = \frac{1}{2} \Gamma_\text{cusp} + \mathcal{O}(\epsilon)$, where $\Gamma_\text{cusp}$ is the cusp anomalous dimension
\begin{align}
\Gamma_\text{cusp} = 4 g^2 + \mathcal{O}(g^4) \, .
\end{align}

The BDS-like normalization contrasts with BDS normalization by the inclusion of a dual conformally invariant function $Y_n$, chosen such that the BDS-like normalization only depends on two-particle Mandelstam invariants,
\begin{align}\label{BDS-like-Norm}
\begin{split}
A_n^{\text{BDS-like}} &=A_n^{\text{BDS}} \exp \left[ \frac{\Gamma_\text{cusp}}{4} Y_n  \right]   \, ,  ~~~ 4 \not|~  n \, , \\
Y_n &= - F_n - 4 A_\text{BDS-like} + \frac{n \pi^2}{4}\, ,
\end{split}
\end{align}
where $F_n $ is (in our conventions) twice the function in Eq. (4.57) of \cite{Bern:2005iz} (one can use an equivalent representation from \cite{Golden:2019kks}) and $A_\text{BDS-like}$ is given on page 57 of \cite{Alday:2010vh}. Since $A_n^{\text{BDS-like}}$ only depends on two-particle Mandelstam invariants, which can be written entirely in terms of frozen variables of the cluster algebra, the BDS-like normalization has the nice feature of not spoiling any cluster adjacency properties. At the same time it means that BDS-like normalized amplitudes will satisfy Steinmann relations \cite{Steinmann:1960on, Steinmann:1960tw, Cahill:1973qp}
\begin{align}
\left.
\begin{array}{rl}
\text{Disc}_{x_{i+1,j}^2}  \left[ \text{Disc}_{x_{i+1 , i+p}^2} (A_n)  \right] &= 0 \, , \vspace{0.05in} \\
\text{Disc}_{x_{i+1,i+p}^2}  \left[ \text{Disc}_{x_{i+1 , j+p+q}^2} (A_n)  \right] &= 0 \, ,
\end{array} \right\} ~~~ 0 < j-i \leq p ~~ \text{or} ~~ q< i-j \leq p+q	 \, .
\end{align}

\subsection{NMHV Amplitudes}

The one-loop $n$-point NMHV amplitude is related to the MHV amplitude by 
\begin{equation}\label{NMHV-MHV}
A^{\text{NMHV}}_n=\mathcal{P}_n A^{\text{MHV}}_n
\end{equation} 
where $\mathcal{P}_n$ is the NMHV ratio function, which at one-loop can be written in the dual conformally invariant form \cite{Drummond:2008bq,Elvang:2009ya}
\begin{align} \label{nptratio}
\mathcal{P}^{(1)}_n = \mathcal{V}_\text{tot} R_{\text{tot}} + \mathcal{V}_{14n} R_{14n} + \sum_{s = 5}^{n-2} \sum_{t= s+2}^{n} \mathcal{V}_{1st} R_{1st} + \text{cyclic} \, .
\end{align}
The transcendental functions $\mathcal{V}_{\text{tot}}$, $\mathcal{V}_{14n}$, and $\mathcal{V}_{1st}$ are given explicitly in Appendix \ref{nmhvfuns}. The function $R_{\text{tot}}$ is given in terms of R-invariants \cite{Drummond:2008vq}
\begin{align}
R_{\text{tot}} = \sum_{s= 3}^{n-2} \sum_{t = s+2}^{n} R_{1 st}\, ,
\end{align}
and $R_{rst}$ are the five-brackets \cite{Mason:2009qx} written in terms of momentum supertwistors as
\begin{align}  \label{fivebracket}
\begin{split}
R_{rst} &= [r , s- 1, s, t-  1 , t]  \, , \\
[a,b,c,d,e] &=  \frac{ \delta^{(4)}(\chi_a \langle b ,c, d, e \rangle + \text{cyclic} )}{ \langle a,b,c,d \rangle \langle b,c,d,e \rangle \langle c,d,e,a \rangle \langle d,e,a,b \rangle \langle e,a,b,c \rangle } \, .
\end{split}
\end{align}
These are special cases of  Yangian invariants \cite{Drummond:2008vq, Drummond:2009fd} and we will henceforth refer to them as such. 

The NMHV BDS-like normalized amplitude is obtained by taking the ratio of equations (\ref{NMHV-MHV}) and (\ref{BDS-like-Norm}), which at one-loop gives
\begin{equation}
A^{(1)\text{ NMHV, BDS-like}}_n=\mathcal{P}^{(1)}_n-\mathcal{P}^{(0)}_nY_n.
\end{equation}

\section{Cluster Adjacency of One-Loop NMHV Amplitudes \label{sec:clustadj}}

In this section we will describe the method we used to test the conjectures in Section \ref{sec:two} and our results.

\subsection{The Symbol and Steinmann Cluster Adjacency}

To compute the symbol of a transcendental function, we follow \cite{Goncharov:2010jf} (see also \cite{Goncharov:2002}).
Only weight two polylogarithms appear at one loop so it is sufficient for us to use the symbols
\begin{align} \label{symbcompute}
\mathcal{S} (\log (R_1) \log(R_2) ) = R_1 \otimes R_2 + R_2 \otimes R_1 \, , ~~~ \mathcal{S} (\text{Li}_2 (R_1) ) = - (1- R_1) \otimes R_1 \, .
\end{align}
Once the symbol of an amplitude is computed, we expand out any cross ratios using \eqref{gencrossratio} and \eqref{xfourbracket}, and perform the bracket test to adjacent symbol entries. It is straightforward to compute the symbol of the expressions in Appendix \ref{nmhvfuns} using \eqref{symbcompute} and we find that the symbol of each of the transcendental functions of \eqref{nptratio}, $\mathcal{V}_{14n}$, $\mathcal{V}_{1, s,t}$, and $\mathcal{V}_\text{tot}$, satisfy Steinmann cluster adjacency (after dropping spurious terms that cancel when expanded out). The function $Y_n$ was already determined to satisfy Steinmann cluster adjacency in \cite{Golden:2019kks}, and hence $A^{(1)\text{ NMHV, BDS-like}}_n$ satisfies \emph{Conjecture 1}. 

It's worth noting that the fact that the one-loop ratio function satisfies Steinmann cluster adjacency for all $n$ is quite remarkable, since when $4|n$ the BDS-like normalization isn't defined and therefore we don't have any expectations on the cluster adjacency of the ratio function from the point of view of \emph{Conjecture 1}.

\subsection{Final Entry and Yangian Invariant Cluster Adjacency}

To study \emph{Conjecture 2}, we follow \cite{Drummond:2018dfd} and start with the BDS-like normalized amplitude expanded as a linear combination of Yangian invariants times transcendental functions
\begin{align}
A^{\text{NMHV, BDS-like}}_{n, L} = \sum_{i}  \mathcal{Y}_i f^{(2L)}_{i} \, ,
\end{align}
We seek a representation of this amplitude that satisfies \emph{Conjecture 2}. Using the bracket test \eqref{sklyaninnorm}, we determine which final symbol entries are not cluster adjacent to all poles of the Yangian invariant multiplying that term. We then rewrite the non-cluster adjacent combinations of Yangian invariants and final entries by using the identities \cite{Mason:2009qx}
\begin{align} \label{fbidens}
[a,b,c,d,e] - [a,b,c,d,f] + [a,b,c,e,f] - [a,b,d,e,f] + [a,c,d,e,f] - [b,c,d,e,f] = 0\, .
\end{align}
until
we are able to reach a form that satisfies final entry cluster adjacency. Note that rewriting in this manner makes the integrability of the symbol no longer manifest. The 6- and 7-point cases were studied in \cite{Drummond:2018dfd}. We checked that this conjecture is true in the 9-point case as well. To get a flavor for our 9-point calculation, consider the following term that we encounter,
which does not manifestly satisfy final entry cluster adjacency:
\begin{align} \label{badcp1}
\begin{split}
&-\frac{1}{2} \left( [1, 2, 3, 4, 5] + [1, 2, 3, 5, 6] + [1, 2, 3, 6, 7] - [1, 2, 4, 5, 7] - [1, 2, 5, 6, 7] \right. \\
&+ \left. [1, 3, 4, 5, 6]  + [1, 3, 4, 6, 7] + [1, 4, 5, 6, 7] - [2, 3, 4, 5, 7] - [2, 3, 5, 6, 7] \right) \\
&\times \left(\log \left(\frac{\langle 1,2,3,4\rangle  \langle 1,7,8,9\rangle
   }{\langle 1,2,7,8\rangle  \langle 1,3,4,9 \rangle }\right) \otimes \langle 3,4,7,8\rangle \right) \, .
\end{split}
\end{align}
To get rid of the non-cluster adjacent combinations of Yangian invariants and final entries, we list all identities \eqref{fbidens} and note that there are 14 cyclic classes of Yangian invariants at 9-points. A cyclic class is generated by taking a five-bracket and shifting all indices cyclically. This collection forms a cyclic class. Solving the identities \eqref{fbidens} for 7 of the 14 cyclic classes in Mathematica (yielding ${{14}\choose{7}} = 3432$ different solutions), we find that at least one solution, for each final entry, brings the symbol to a final entry cluster adjacent form. For the example \eqref{badcp1}, one of the combinations from these solutions, that is cluster adjacent, takes the form
\begin{align}
\begin{split}
&- \frac{1}{2} ([1, 2, 3, 4, 8] - [1, 2, 3, 7, 8] + [1, 2, 4, 7, 8] -  [1, 3, 4, 7, 8] \\
&+ [2, 3, 4, 7, 8] + [3, 4, 5, 6, 7]) \left(\log \left(\frac{\langle 1,2,3,4\rangle  \langle 1,7,8,9\rangle }{\langle 1,2,7,8\rangle  \langle 1,3,4,9 \rangle }\right) \otimes \langle 3,4,7,8\rangle \right) \, .
\end{split}
\end{align}
One can check that the complete set of Yangian invariants that are cluster adjacent to $\langle 3,4,7,8 \rangle$ is given by
\begin{align} \label{goodfb1}
\begin{split}
&\{[1, 2, 3, 4, 7], [1, 2, 3, 4, 8], [1, 2, 3, 4, 9], [1, 2, 3, 7, 8], [1, 2, 3, 7, 9], [1, 2, 3, 8, 9], \\
&[1, 2, 4, 7, 8], [1, 2, 4, 7, 9], [1, 2, 4, 8, 9], [1, 2, 7, 8, 9], [1, 3, 4, 7, 8], [1, 3, 4, 7, 9], \\
&[1, 3, 4, 8, 9], [1, 3, 7, 8, 9], [1, 4, 7, 8, 9], [2, 3, 4, 7, 8], [2, 3, 4, 7, 9], [2, 3, 4, 8, 9], \\
& [2, 3, 7, 8, 9], [2, 4, 7, 8, 9], [3, 4, 5, 6, 7], [3, 4, 5, 6, 8], [3, 4, 5, 7, 8], [3, 4, 6, 7, 8],  \\
&[3, 4, 7, 8, 9], [3, 5, 6, 7, 8], [4, 5, 6, 7, 8]\}\, .
\end{split}
\end{align}
At 10-points this method becomes much more computationally intensive as we have 26 cyclic classes. If we follow the same procedure as for 9-points, we would have to check cluster adjacency of ${{26}\choose{13}} = 10400600$ solutions per final entry with non cluster adjacent Yangian invariants.

\section{Cluster Adjacency and Weak Separation \label{sec:weaksep}}

In our study of one-loop NMHV amplitudes we observed some general cluster adjacency properties of symbol entries and Yangian invariants involved in the one-loop NMHV amplitude.  Let us denote the various types of symbol letters by
\begin{align}
a_{1; i j} &= \langle i-1 , i , j-1 , j \rangle \, , \label{pluck1}\\
\begin{split}
a_{2; ijk} &= \langle i (j \, j+1) (k \, k+1) (i-1 \, i+1) \rangle \\
&= \langle i , j , j+1 , i-1 \rangle \langle i , k , k+1 , i+1 \rangle -  \langle i , j , j+1 , i+1 \rangle \langle i , k , k+1 , i-1 \rangle \, , \label{pluck2}
\end{split} \\
\begin{split}
a_{3 ; ijkl} &= \langle i (j \, j+1) (k \, k+1) (l \, l+1) \rangle \\
&= \langle i ,  j , k , k+1 \rangle \langle i , j+1 , l , l+1 \rangle -\langle i ,  j+1 , k , k+1 \rangle \langle i , j , l , l+1 \rangle   \,.  \label{pluck3}
\end{split}
\end{align}
In this section we summarize their cluster adjacency properties as determined by the \emph{bracket test}.

\begin{figure}[t!]
\centering
\includegraphics[width=0.5\textwidth]{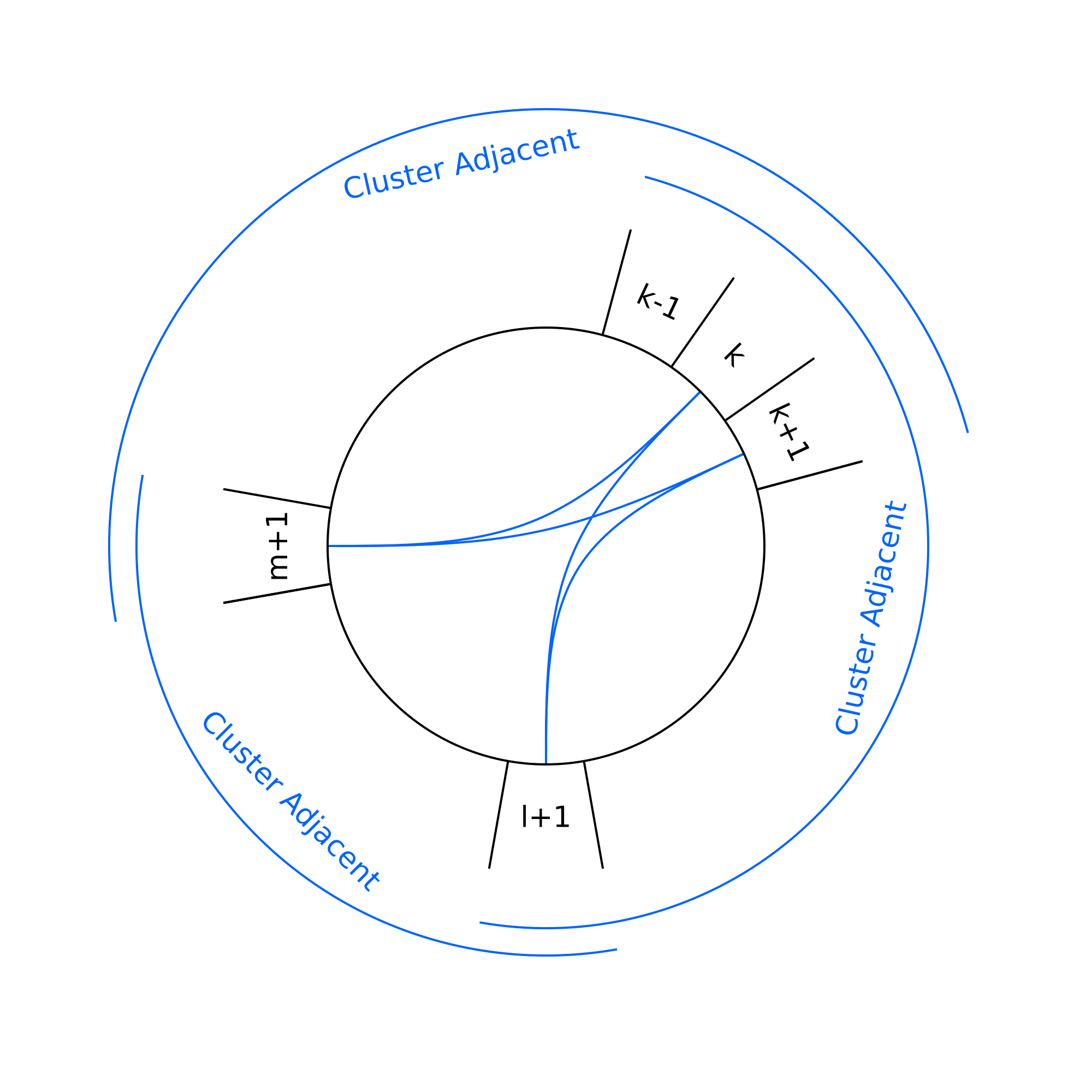}
\captionsetup{justification=justified, margin = 25pt}
\caption{Weak separation graph indicating that if both $i$ and $j$ are within any of the blue regions (or on the blue chords), then $\langle i-1 , i , j-1 , j \rangle$ is cluster adjacent to $\langle k (l \, l+1) (m \, m+1) (k-1 \, k+1) \rangle$. }
\label{fig:steinmannplucker1}
\end{figure}

\begin{figure}[t!]
\centering
\includegraphics[width=0.5\textwidth]{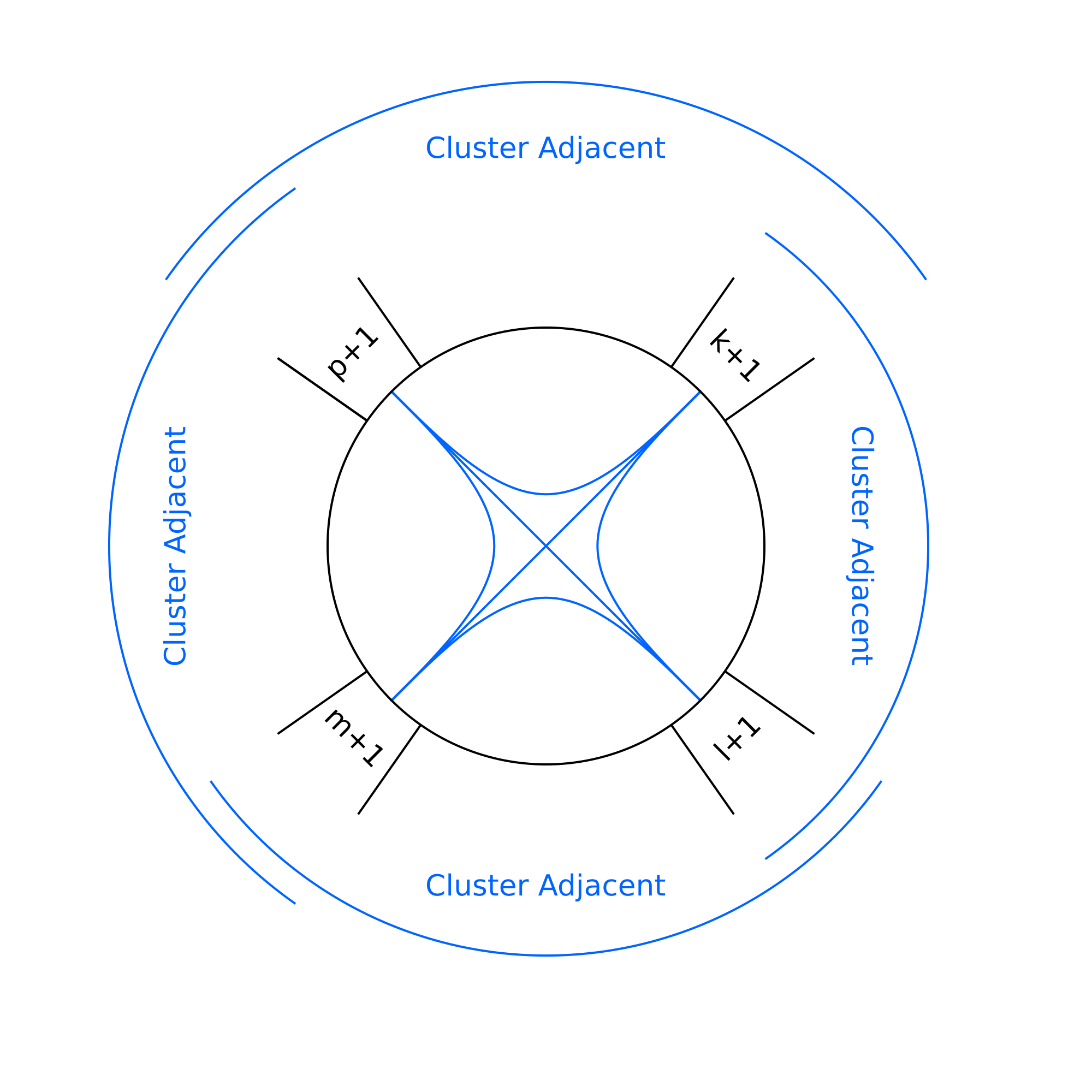}
\captionsetup{justification=justified, margin = 25pt}
\caption{Weak separation graph indicating that if both $i$ and $j$ are within any of the blue regions (or on the blue chords), then $\langle i-1 , i , j-1 , j \rangle$ is cluster adjacent to $\langle k (l \, l+1) (m \, m+1) (p \, p+1) \rangle$. }
\label{fig:steinmannplucker2}
\end{figure}

First consider $a_{1;i j}$ and $a_{2; k l m}$. We observe that these variables are adjacent if they satisfy a generalized notion of weak separation \cite{LZ,Oh:2011}. In particular we find that
\begin{align}
\begin{split}
&\langle i-1 , i , j-1 , j \rangle ~~ \text{and} ~~ \langle k (l \, l+1) (m \, m+1) (k-1 \, k+1) \rangle  ~~ \text{are cluster adjacent iff} \\
&\{ i, j \} \in \{ k+1 ,\ldots , l+1 \}  \lor\{ i, j \} \in  \{ l+1 , \ldots,  m+1 \} \lor \{ i, j \} \in \{ m+1, \ldots , k \} ~~ \text{or} \\
&\{i = k, j = l+1 \}  \lor \{ i = k , j=m+1 \} \lor \{ i = k+1 , j = l+1 \} \lor \{ i = k+1 , j= m+1 \} \, .
\end{split}
\end{align}
This adjacency statement can be represented by drawing a circle with labeled points $\{ 1 , \ldots, n\} $ appearing in cyclic order, as in Figure \ref{fig:steinmannplucker1}.
For the variables $a_{1; ij}$ and $a_{3; k lm p}$ we observe
\begin{align}
\begin{split}
&\langle i-1 , i , j -1 , j \rangle ~~ \text{and} ~~ \langle k (l \, l+1) (m \, m+1) (p \, p+1 ) \rangle  ~~ \text{are cluster adjacent iff} \\
& \{ i, j \} \in \{ k+1, \ldots ,l+1\} \lor \{ i, j \} \in \{ l+1, \ldots , m+1 \}  \lor \{ i, j \} \in \{ m+1 ,\ldots , p+1 \}\\
& \lor \{ i, j \} \in \{ p+1 , \ldots,  k+1 \} ~~ \text{ or} ~~ \{ i = k +1 , j = l+1\}  \lor  \{ i = l+1 , j = m+1\} \\
& \lor  \{ i = m+1 , j = p+1\} \lor \{ i = p+1, j = k+1\}  \lor \{ i = k+1 , j = m+1 \}  \\
& \lor \{ i = l+1 , j = p+1 \}.
\end{split}
\end{align}
This statement is represented in Figure \ref{fig:steinmannplucker2}.

For Pl{\"u}cker coordinate of type \eqref{pluck1} and Yangian invariants \eqref{fivebracket}, we observe
\begin{align}
\begin{split}
&\langle i-1 , i , j -1 , j \rangle ~~ \text{and} ~~ [a,b,c,d,e]~~ \text{are cluster adjacent iff} \\
&\{ a,b,c,d,e \} \subset {{\{ i-1 ,i , \ldots, j-1 , j\}}\choose{5}} \cup {{\{j-1 , j, \ldots , i-1 , i \}}\choose{5}}.
\end{split}
\end{align}
Next up the variables \eqref{pluck2} and Yangian invariants \eqref{fivebracket} are observed to have the adjacency condition
\begin{align}
\begin{split}
&\langle i (j \, j+1 ) (k \, k+1) (i-1 \, i+1) \rangle ~~ \text{and} ~~ [a,b,c,d,e]~~ \text{are cluster adjacent iff} \\
&\{ a,b,c,d,e \} \subset \{ \{ i, j,j+1,k,k+1 \} \} \cup {\{ i, i+1 , \ldots, j, j+1 \}\choose{5}} \\
&\cup  {\{ j, j+1 , \ldots, k, k+1 \}\choose{5}}  \cup  {\{ k, k+1 , \ldots, i-1, i \}\choose{5}}.
\end{split}
\end{align}
Finally, for variables \eqref{pluck3} and Yangian invariants \eqref{fivebracket}, we observe adjacency when
\begin{align}
\begin{split}
&\langle i (j \, j+1 ) (k \, k+1) (l \, l+1) \rangle ~~ \text{and} ~~ [a,b,c,d,e]~~ \text{are cluster adjacent iff} \\
&\{ a,b,c,d,e \} \subset {{\{ i , \ldots, j , j+1 \}}\choose{5}} \cup {{\{ i , j , j+1 , \ldots, k , k+1 \}}\choose{5}} \\
&\cup {{\{ i, k, k+1 , \ldots, l , l+1 \}}\choose{5}} \cup {{\{ l , l+1 , \ldots, i \}}\choose{5}}  \, .
\end{split}
\end{align}
The statements about cluster adjacency in this section hint at a generalization of the notion of weak separation for Pl{\"u}cker coordinates \cite{LZ,Oh:2011}. We are only able to verify these statements ``experimentally'' via the bracket test. To prove such statements, we look to Theorem 1.6 of \cite{Oh:2011} which states that: given a subset $\mathcal{C}$ of ${{\{ 1, \ldots, n\}}\choose{4}}$, the set of Pl{\"u}cker coordinates $\{ p_I \}_{I \in \mathcal{C}}$ forms a cluster in the $\Gr(4,n)$ cluster algebra iff $\mathcal{C}$ is a maximally weakly separated collection. Maximally weakly separated means that if $\mathcal{C} \subseteq {{\{ 1, \ldots, n\}}\choose{4}}$ is a collection of pairwise weakly separated sets and $\mathcal{C}$ is not contained in any larger set of of pairwise weakly separated sets, then the collection $\mathcal{C}$ is maximally weakly separated. To prove the cluster adjacency statements made in this section, we would have to prove that there exists a maximally weakly separated collection containing all the weakly separated sets proposed in for each pair of coordinates/Yangian invariants considered in this section. We leave this to future work.

\acknowledgments

We are grateful to \"O.~G\"urdo\u{g}an and M.~Parisi for stimulating discussions and for comments on the draft.
This work was supported in part by the US Department of Energy under contract {DE}-{SC}0010010 Task A and by Simons Investigator Award \#376208 (AV).

\appendix

\section{$n$-point NMHV Transcendental Functions \label{nmhvfuns}}

In this Appendix we present the transcendental functions contributing to the NMHV ratio function \eqref{nptratio}, from \cite{Elvang:2009ya}. All functions are written in a dual conformally invariant form, in terms of cross ratios
\begin{align} \label{gencrossratio}
u_{ijkl} = \frac{x_{ik}^2 x_{jl}^2}{x_{il}^2 x_{jk}^2}
\end{align}
of dual momenta \eqref{xfourbracket}. The functions $\mathcal{V}_{1,s,t}$ are given by
\begin{align} \label{v1st}
\begin{split}
\mathcal{V}_{1,s,t} &= \text{Li}_2 (1 - u_{12t4} ) - \text{Li}_2 (1- u_{12 ts}) + \sum_{i=5}^s \left[ \text{Li}_2 (1- u_{2, i+2, i-1, i} ) - \text{Li}_2 (1- u_{12 i , i-1})  \right. \\
&- \text{Li}_2 (1- u_{1,i+2 , i -1, i}) - \frac{1}{2} \ln (u_{2 1 i, i+2} ) \ln (u_{1, i+2, i-1, i} ) - \frac{1}{2} \ln (u_{12ti} ) \ln (u_{1t, i-1, i})  \\
&- \frac{1}{2} \left. \ln (u_{2, i-1, t, i+2} ) \ln (u_{12 i, i-1}) \right] \, , ~~~~~ \text{for}~~~ 5 \leq s\, , t \leq n-1 \, ,
\end{split}
\end{align}
where $5 \leq s \leq n-2$ and $s+2 \leq t \leq n$, and
\begin{align} \label{v1sn}
\begin{split}
\mathcal{V}_{1,s,n} &= \text{Li}_2 (1- u_{2 s n, n-1}) + \text{Li}_2 (1- u_{214, n-1}) + \ln (u_{2 s n , n-1} ) \ln (u_{21s,n-1}) \\
&+ \sum_{i =5}^s \left[ \text{Li}_2 ( 1- u_{2,i+2,i-1, i}) - \text{Li}_2 (1- u_{12i,i-1}) - \text{Li}_2 (1- u_{1,i+2 , i-1, i} ) \right. \\
&- \frac{1}{2} \ln (u_{21 i , i+2} ) \ln (u_{1,i+2, i-1, i})  -\frac{1}{2} \ln (u_{12,n-1, i}) \ln (u_{1,n-1, i-1,i}) \\
&-  \left.  \frac{1}{2} \ln (u_{2,i-1,n-1,i+2} ) \ln (u_{12i,i-1})  \right] - \frac{\pi^2}{6}, ~~~ \text{for} ~~~ 4 \leq s \leq n-3 \, ,
\end{split}
\end{align}
where the sum empty sum is understood to vanish for $s = 4$. The function $\mathcal{V}_{1,n-2,n}$ is given by
\begin{align} \label{v1n2n}
\begin{split}
\mathcal{V}_{1,n-2,n} &= \text{Li}_2 (1- u_{2n , n-3, n-2} ) - \text{Li}_2 (1- u_{12,n-2,n-3}) + \text{Li}_2 (1- u_{2,n-3,n,n-1} ) \\
&+ \text{Li}_2 (1- u_{214, n-1}) - \ln (u_{n1,n-3,n-2}) \ln \left( \frac{u_{12,n-2,n-1}}{u_{2,n-3, n-1,n} }  \right)  \\
&+ \ln (u_{2,n-3, n,n-1}) \ln (u_{21, n-3, n-1})  + \sum_{i = 5}^{n-3} \left[ \text{Li}_2 (1- u_{2,i+2,i-1,i} )  \right. \\
&- \text{Li}_2 (1 - u_{12i,i-1}) - \text{Li}_2 (1- u_{1,i+2,i-1,i}) -  \frac{1}{2} \ln (u_{21i,i+2}) \ln (u_{1,i+2,i-1,i})     \\
&-  \frac{1}{2} \ln (u_{12,n-1,i}) \ln (u_{1,n-1,i-1,i}) - \left. \frac{1}{2} \ln (u_{2,i-1,n-1,i+2} ) \ln (u_{12i,i-1})   \right] - \frac{\pi^2}{6} \, .
\end{split}
\end{align}
Finally $\mathcal{V}_\text{tot}$ is given by two different formulas, one for $n = 8$ and one for $n>8$. For $n=8$ we have
\begin{align} \label{vtot8}
8 \mathcal{V}_\text{tot}^{n=8} = - \text{Li}_2 (1- u_{1247}^{-1}) + \frac{1}{2} \sum_{i =4}^6 \text{Li}_2 (1- u_{12i, i+1}^{-1} ) + \frac{1}{4} \ln (u_{8145} ) \ln \left( \frac{u_{1256} u_{3478}}{u_{2367} } \right) + \text{cyclic} \, ,
\end{align}
while for $n>8$ we have
\begin{align} \label{vtot}
\begin{split}
n \mathcal{V}_{\text{tot}} &= - \text{Li}_2 (1- u_{124,n-1}^{-1} )+ \frac{1}{2} \sum_{i=4}^{n-2} \text{Li}_2 (1- u_{12 i , i+1}^{-1}) \\
&+ \frac{1}{2} \ln (u_{n 1 3 4} ) \ln (u_{136, n-2} ) - \frac{1}{2} \ln (u_{n 145} ) \ln (u_{236,n-2} u_{2367}) + v_n + \text{cyclic} \, ,
\end{split}
\end{align}
where
\begin{align}
n~\text{odd}  &: ~~~ v_n = \sum_{i=4}^{\frac{n-1}{2}} \ln (u_{n1 i , i +1} ) \sum_{j=1}^{i-1} \ln(u_{j, j+1, i+j, n-i+j} ) \, , \\
n~\text{even}  &: ~~~ v_n = \sum_{i=4}^{\frac{n-1}{2}} \ln (u_{n1 i , i +1} ) \sum_{j=1}^{i-1} \ln(u_{j, j+1, i+j, n-i+j} ) + \frac{1}{4} \ln (u_{n1, \frac{n}{2} , \frac{n}{2} +1} ) \sum_{i=1}^{\frac{n-2}{2}} \ln (u_{i, i+1, i+\frac{n}{2}, i+ \frac{n}{2} +1} )\, .
\end{align}

\end{document}